\documentclass[12pt]{article}

\usepackage{epsfig,rotating}   

\newcommand{\pcc}{cm$^{-3}$}
\newcommand{\pccps}{cm$^{-3}$s$^{-1}$}
\newcommand{\htwosofour}{H$_2$SO$_4$}
\newcommand{\nhthree}{NH$_3$}
\newcommand{\htwoo}{H$_2$O}

\newcommand{\lappeq}{\mathrel{\rlap{\raise.5ex\hbox{$<$}}{\lower.5ex\hbox{$\sim$}}}}
\newcommand{\gappeq}{\mathrel{\rlap{\raise.5ex\hbox{$>$}}{\lower.5ex\hbox{$\sim$}}}}

\begin{document}           

\pagestyle{empty}

\begin{titlepage}

\begin{center}

EUROPEAN ORGANIZATION FOR NUCLEAR RESEARCH

{\small
\begin{tabbing}
 \=  \hspace{117mm}  \=  \kill 
 \>  \>CERN/SPSC 2000-030 \\
 \>  \>SPSC/P317 Add.1  \\ 
 \>  \>August 4, 2000     \\  
\end{tabbing}  }

\vspace{-30mm}
\begin{figure}[htbp]
  \begin{center}
      \makebox{\epsfig{file=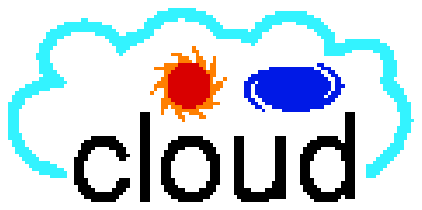,height=18mm} \hspace{0.2mm}}
  \end{center}  
\end{figure}

\textbf{ADDENDUM TO THE CLOUD PROPOSAL} \\[2ex] 

 {\footnotesize
 B.\,Fastrup, E.\,Pedersen \\ 
 \emph{University of Aarhus, Institute of Physics and Astronomy, 
 Aarhus, Denmark}  \\[1ex] 
 E.\,Lillestol, E.\,Thorn  \\
 \emph{University of Bergen, Institute of Physics, Bergen,  Norway}  
 \\[1ex]
 M.\,Bosteels, A.\,Gonidec, J.\,Kirkby*, 
 S.\,Mele, P.\,Minginette, B.\,Nicquevert, 
 D.\,Schinzel, W.\,Seidl \\ 
 \emph{CERN, Geneva, Switzerland} \\[1ex] 
 P.\,Grunds\mbox{\o}e, N.\,Marsh, J.\,Polny, H.\,Svensmark  \\
 \emph{Danish Space Research Institute, Copenhagen, Denmark} \\[1ex]  
  Y.\,Viisanen  \\
 \emph{Finnish Meteorological Institute, Helsinki,
 Finland} \\[1ex]
 K.\,Kurvinen, R.\,Orava  \\
 \emph{University of Helsinki, Institute of Physics,  
 Helsinki, Finland}  \\[1ex] 
 K.\,H\"{a}meri, M.\,Kulmala, L.\,Laakso, J.M.\,M\"{a}kel\"{a}, 
 C.D.\,O'Dowd \\ 
 \emph{University of Helsinki, Lab.\,of Aerosol and
 Environmental Physics, Helsinki, Finland} \\[1ex]
 V.\,Afrosimov, A.\,Basalaev, M.\,Panov  \\ 
 \emph{Ioffe Physical Technical Institute, Dept.\,of Fusion
 Technology, St.\,Petersburg, Russia} \\[1ex]
 A.\,Laaksonen, J.\,Joutsensaari  \\
 \emph{University of Kuopio, Department of Applied Physics, Kuopio,
 Finland}  \\[1ex] 
 V.\,Ermakov, V.\,Makhmutov, O.\,Maksumov, P.\,Pokrevsky,
 Y.\,Stozhkov, N.\,Svirzhevsky  \\
 \emph{Lebedev Physical Institute, Solar and Cosmic Ray Research
 Laboratory, Moscow, Russia} \\[1ex] 
 K.\,Carslaw, Y.\,Yin  \\
 \emph{University of Leeds, School of the Environment, Leeds, United
 Kingdom} \\[1ex]
 T.\,Trautmann  \\
 \emph{University of Mainz, Institute for Atmospheric Physics,
  Mainz, Germany} \\[1ex]
 F.\,Arnold, K.-H.\,Wohlfrom  \\
 \emph{Max-Planck Institute for Nuclear Physics (MPIK), Atmospheric
Physics Division,
 Heidelberg, Germany} \\[1ex]
 D.\,Hagen, J.\,Schmitt, P.\,Whitefield  \\
 \emph{University of Missouri-Rolla, Cloud and Aerosol Sciences
 Laboratory,  Rolla, USA} \\[1ex] 
 K.\,Aplin, R.G.\,Harrison  \\
 \emph{University of Reading, Department of Meteorology,
 Reading, United Kingdom} \\[1ex]
 R.\,Bingham, F.\,Close, C.\,Gibbins, A.\,Irving, B.\,Kellett,
M.\,Lockwood \\
 \emph{Rutherford Appleton Laboratory, 
 Space Science \& Particle Physics Depts., Chilton, United
 Kingdom} \\[1ex] 
 D.\,Petersen, W.W.\,Szymanski, P.E.\,Wagner, A.\,Vrtala  \\ 
 \emph{University of Vienna, Institute for Experimental Physics,
 Vienna, Austria}  \\[1ex]   }                        
 {\normalsize CLOUD$^\dagger$  Collaboration} \\
\end{center}
\vspace{-5mm}

\vfill

\noindent \rule{60mm}{0.1mm} \\ {\footnotesize
$*$) spokesperson  \\
$\dagger$) Cosmics Leaving OUtdoor Droplets }
\date{}

\end{titlepage}

\pagestyle{plain}     

\pagenumbering{roman}  
\setcounter{page}{2}  
\newpage \tableofcontents 

 
\newpage
\pagestyle{plain}     
\pagenumbering{arabic}  
\setcounter{page}{1}  

%
%
\section{Introduction}  \label{sec_introduction}

\subsection{Overview}  \label{sec_overview}

Following the request of the CERN SPSC at its meeting on 23 May 2000,
this report provides further details on the experimental programme and
beam schedule of CLOUD.  Specifically, we describe  (in \S
\ref{sec_cn_nucleation} and \S \ref{sec_cn_growth}) two examples of the
type of experiments to be performed by CLOUD, and their expected
precisions. The experimental and theoretical motivation for these
experiments is summarised in \S \ref{sec_motivation}.  The beam
requirements and schedule are elaborated in \S
\ref{sec_accelerator_requirements}.   We also summarise the progress on
the detector design since the  proposal \cite{cloud_proposal} in
\S \ref{sec_progress}, and the expected scope of the aerosol and cloud
simulations for the experiment in \S \ref{sec_simulations_summary}.

\subsection{Choice of experiments}  \label{sec_choice}

The ions and radicals produced in the atmosphere by galactic cosmic rays
(GCRs) may influence cloud microphysics in several ways:

\begin{enumerate}
\item Nucleation (creation) of new condensation nuclei (CN) from trace
condensable vapours.
\item Growth of CN into cloud condensation nuclei (CCN), which seed
cloud droplets.
\item Activation of CCN into cloud droplets in the presence of
supersaturated water vapour.
\item Creation of ice nuclei, which cause the freezing of supercooled
liquid droplets in clouds.
\end{enumerate}

If the rate or efficiency of any of these processes is significantly
modified by ions in the atmosphere, then the GCR intensity could modify 
the droplet number concentrations of clouds and thereby affect their
lifetimes and radiative properties.  Motivated by the experimental and
theoretical considerations summarised in \S
\ref{sec_motivation}, we have selected for this report one example
experiment from each of the first two processes listed above,
specifically:
\begin{description}
\item[Exp.1:] Binary homogeneous nucleation of ultrafine condensation
nuclei (UCN) of sulphuric acid and water, under atmospheric conditions
(\S
\ref{sec_cn_nucleation}).
\item[Exp.2:] Growth of these UCN into CCN in the presence of additional
vapours: sulphuric acid, water, ammonia and low volatility organic
compounds (LVOCs),
 again under atmospheric conditions (\S \ref{sec_cn_growth}).  
\end{description} 
 These examples are part of the category I experiments (``Creation
\& growth of aerosols'') listed in Table 3 (p.29) of the CLOUD proposal.

\section{Progress on the detector design} 
\label{sec_progress}

\subsection{Summary}  \label{sec_summary}

\begin{figure}[hbp]
  \begin{center}
      \makebox{\epsfig{file= 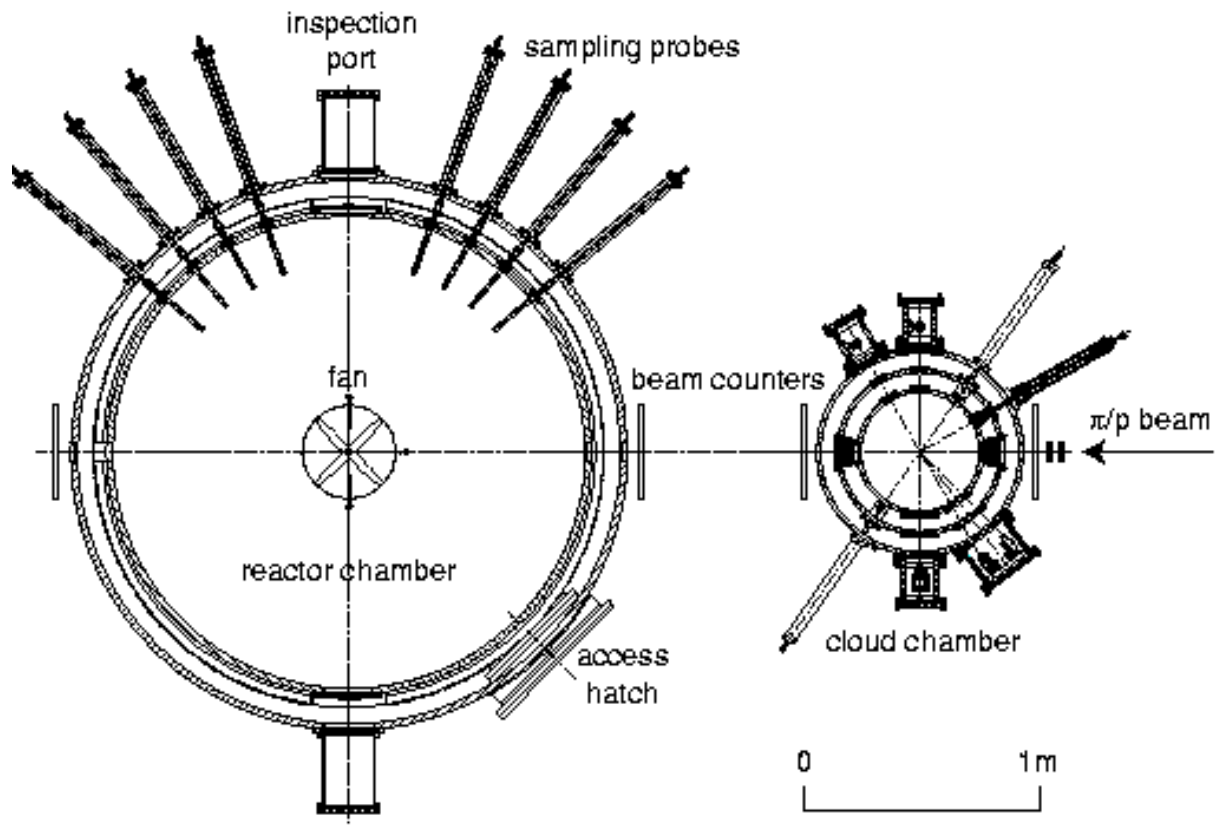,width=160mm}}
  \end{center}
  \vspace{-5mm}
  \caption{Horizontal section of the cloud chamber and the new reactor
chamber.}
  \label{fig_reactor_chamber_h}
\end{figure}

\begin{sidewaysfigure}[htbp]
  \begin{center}
      \makebox{\epsfig{file=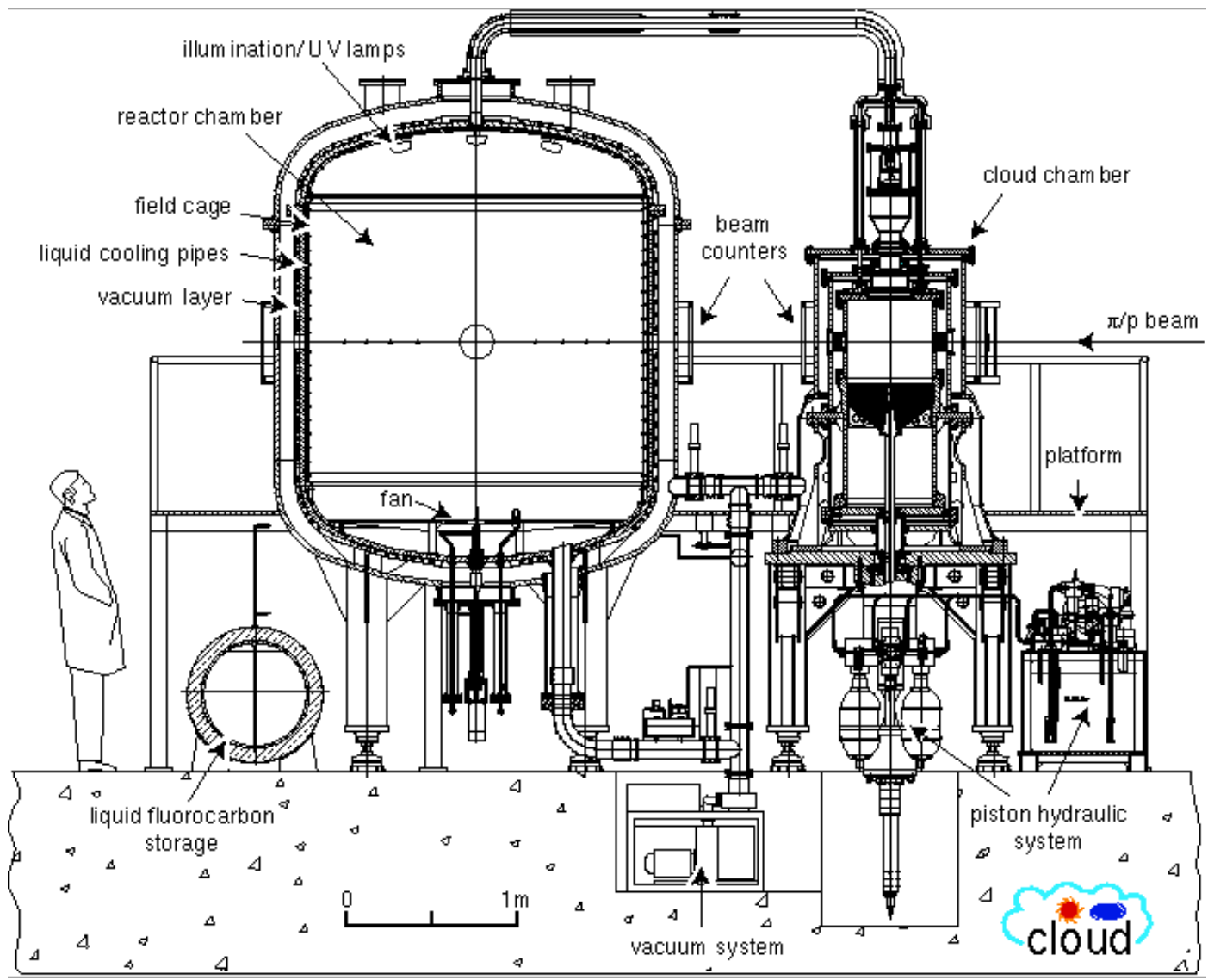,height=150mm}}
  \end{center}
  \caption{Vertical section showing the cloud chamber and the new
reactor chamber.   The latter combines the functions of the original
buffer expansion tank and flow reactor chamber, and replaces them.}
  \label{fig_reactor_chamber_v}    
\end{sidewaysfigure}

Since the proposal we have improved the detector design by replacing the
buffer expansion tank and flow chamber (described in \S 4 of the
proposal) by a single device, a
\emph{reactor chamber} (Figs.\,\ref{fig_reactor_chamber_h} and
\ref{fig_reactor_chamber_v}).  

The reactor chamber will be located in the beamline, immediately behind
the cloud chamber.   It performs the same functions as the devices it
replaces, but with improved performance, namely:
\begin{enumerate}
\item Buffer tank for small expansions.  This functions in the same way
as described previously in the proposal, providing a second method to
produce small expansions.
\item A reactor chamber for providing the source of reacted gas/aerosols
for the external detectors.  This replaces the function of the original
flow chamber but with the improvement that it is now effective for long
growth-time experiments extending for several days, since diffusion
losses to the walls are strongly reduced.  Since the aerosol and trace
gas lifetimes from wall losses should scale as the linear dimension of
the vessel (i.e. the ratio of its volume to surface area), we expect the
reactor chamber will have about a factor four longer particle lifetime
than the cloud chamber.   The cloud chamber is suitable for growth
experiments lasting up to about a day.  For longer growth-time
experiments the reactor chamber can now  provide samples of reacted
gas/aerosols for analysis in the cloud chamber itself since  its volume 
is much larger than the cloud chamber (by a factor 64).
\end{enumerate} 

The estimated extra cost of the reactor chamber over the original buffer
expansion tank and flow chamber it replaces is 30 kCHF.

\subsection{Reactor chamber design}  
\label{sec_reactor_chamber_design}

The reactor chamber is a cylinder of internal dimensions 2 m (diameter)
$\times$ 2 m (height). The inner vessel can be operated between
atmospheric pressure and a vacuum.  It is constructed from aluminium,
with black teflon lining the inner surface. The teflon is made partly
conductive in order to prevent any charge buildup. An inner field cage
provides a simple clearing field.   

The reactor chamber is exposed to the same beam as the cloud chamber and
operated at the same temperature and pressure.  However, since it simply
provides a container for trace gases and aerosols to react in the
presence of the beam (i.e. no droplet activation is involved), the
requirements on the temperature stability are less demanding than for
the cloud chamber, and a stability of a few $\times$ 0.1\,K is
adequate.  The temperature is controlled with liquid fluorocarbon
circulated in pipes around the outer surface of the reactor vessel.
Heater cables are also wrapped around the reactor vessel to allow
cleaning by  bakeout under vacuum. A simplified thermal insulation is
employed, comprising  superinsulation and a vacuum layer.  The liquid
system is independent of the one used for the cloud chamber in order to
maintain the precise temperature stability of 0.01\,K   required for the
cloud chamber.  

The reactor chamber is equipped with gas/aerosol inlet and outlet
pipes.  A small fan is installed inside the vessel to provide the option
of slow stirring of the gas filling to assist, where necessary,
homogeneous mixing throughout the large volume.  A special  inlet pipe
located near the fan provides fresh trace gas to replace losses to the
walls or to aerosol growth.  The reactor vessel is also equipped with
sampling probes to  extract gas and aerosols for the external detectors 
(mass/ion/mobility spectrometers, trace gas analysers and aerosol
particle sizers).  The tip of each sampling probe can be independently 
adjusted to any radial position ($0 < r < 1$\,m) inside the vessel.  The
reactor vessel also includes ports connecting to the vacuum system and
to the active volume of the cloud chamber.  

Access to the inside of the reactor vessel is provided by a removable
hatch in the side wall and, for full access, the top of the vacuum tank
and reactor vessel are also removable. The vessel is equipped with
internal illumination and two inspection ports.  Internal UV lamps are
also provided for experiments involving photochemical processes.

\subsection{Reactor chamber performance}  
\label{sec_reactor_chamber_performance}

The performance of the reactor chamber principally concerns the lifetime
of trace gases and aerosols due to wall losses.  Aerosols that strike
the walls will be lost due to attractive van der Waals forces.  Since
the diffusion coefficient of aerosol particles is small, the wall losses
due to thermal diffusion are small (see Fig.\,42 of the proposal).   
Trace gas molecules, on the other hand, have a high diffusion rate. 
However these molecules strike the walls at high velocity and mostly   
rebound (which is, after all, the origin of gas pressure in a vessel). 
The fraction that do not rebound depends on temperature and on the
nature of the gas molecules and the surface material. 

The experience of the AIDA aerosol facility \cite{bunz} at
Forschungszentrum  Karlsruhe is instructive in this regard.  AIDA is a
large cylindrical vessel of internal dimensions 4 m (diameter) $\times$
7\,m (height) with an inner ceramic lining. The temperature is
controlled to 0.5\,K precision by cold air circulating around the
outside of the  vessel. It is equipped with an internal fan to ensure
homogeneous mixing.  Operation of the fan is found to have only a small
effect on aerosol lifetimes.   The measured 1/$e$ lifetimes in AIDA for
trace O$_3$ and NO$_2$ gases at room temperature are 70 h and 370 h,
respectively.  Since the lifetimes should scale as the linear dimension
of the vessel, this implies the CLOUD reactor chamber should have
corresponding lifetimes of about 30 h and 150 h, respectively.  In fact,
since teflon is superior to ceramic as a lining material, these lifetime
figures are probably conservative.  During long-duration experiments
with CLOUD, the losses of trace vapours due to wall adhesion or to
aerosol growth will be compensated by a continuous small inflow of trace
gases.

\section{Experimental and theoretical motivation} 
\label{sec_motivation}

\subsection{Present experimental knowledge}  
\label{sec_experimental_knowledge}

There are only sparse experimental data on the effect of ions in the
atmosphere on new particle formation---and none, to our knowledge, on
the effect of ions on particle growth from CN to CCN.  Observations have
been made of nucleation bursts of CN in the atmosphere that cannot be
explained  by classical theories. For example, H\~{o}rrak \emph{et al.}
\cite{horrak} reported the spontaneous formation of bursts of
intermediate size ions in urban air, which they suggest may be due to
ion-induced nucleation.  Also, Clarke
\emph{et al.} \cite{clarke} observed formation of new ultrafine
particles in the tropical marine (Pacific) boundary layer that could not
be explained by classical binary (\htwosofour-\htwoo) homogeneous
nucleation theory at the low ambient concentrations of sulphuric acid
(1--5$ \cdot 10^7$ molecules~cm$^{-3}$).  

 However a recent study  by Yu and Turco \cite{yu00a} based on 
 an ion-mediated model \emph{is}  
 able to reproduce the observations of Clarke \emph{et al.}, as
described below.   Their model indicates that the  nucleation rate of
fresh CN in the marine boundary layer\footnote{The
\emph{boundary layer}   extends from the Earth's surface up to about 1 
km; the region extending from about 1 km to the tropopause is known as
the \emph{free troposphere}.} is generally limited by the available ion
production rate from GCRs.  In contrast, the nucleation rate in the
upper atmosphere is generally limited by the trace vapour concentration
since the temperatures are lower and the trace vapour saturation ratios
correspondingly higher, and so binary homogeneous nucleation can occur
at an appreciable rate.  This could explain why satellite observations
show a solar modulation signal only in clouds below about 3 km (Fig.\,6
in the CLOUD proposal).

Direct experimental evidence that ions are involved in the nucleation of
new particles under atmospheric conditions  is lacking.  However
positive effects  with ions have been seen. For example, Bricard
\emph{et al.}\,\cite{bricard} observed new particle production in
filtered (aerosol-free) Paris air exposed to very high radiation doses
\mbox{($3\cdot10^8$~Bq~m$^{-3}$ $\times$ 300 s).}   Vohra \emph{et
al.}\,\cite{vohra}, on the other hand, carried out experiments with
radon at naturally-occurring ionisation levels of 3--15 Bq m$^{-3}$  and
observed new particle production proportional to ionisation rate, but
using artificial air containing high concentrations of trace gases
(300~ppb SO$_2$, 100 ppb C$_2$H$_4$ and  160~ppb O$_3$).

In short, many of the CLOUD experiments will be the first of their kind 
and, as such, there are good prospects for making important advances in
our understanding of the importance of ionising particle radiation on
aerosol and cloud physics.

\subsection{Physical model of ion-mediated nucleation and growth} 
\label{sec_physical_model}

Despite intensive research over several decades, the origin of the
ubiquitous  background of ultrafine aerosols in the troposphere has not
yet been determined.  Moreover even the fundamental mechanism that leads
to new particle formation remains poorly understood.  It has been
suggested that ionisation from GCRs plays a key role in the formation of
new aerosol particles 
\cite{arnold80}--\cite{yu00b}, and indeed this will be one of the major
areas of study for CLOUD.

The classical theory of binary  \htwosofour-\htwoo\ homogeneous
nucleation fails to explain observations of new ultrafine particle
formation in clean regions of the lower atmosphere, such as occurs over
oceans and  in pristine continental air
\cite{clarke,weber97,weber99,kulmala00}.  Typically the nucleation rates
predicted by classical theory are far  lower (by as much as 10 orders of
magnitude) than the experimentally-observed rates.   Recent modelling
work 
\cite{yu00a} demonstrates that thermodynamically-stable charged
clusters, caused by vapours condensing onto ions produced by GCRs, can
form at much lower ambient  vapour concentrations and grow significantly
faster than neutral clusters.

\begin{sidewaysfigure}
  \begin{center}
      \makebox{\epsfig{file=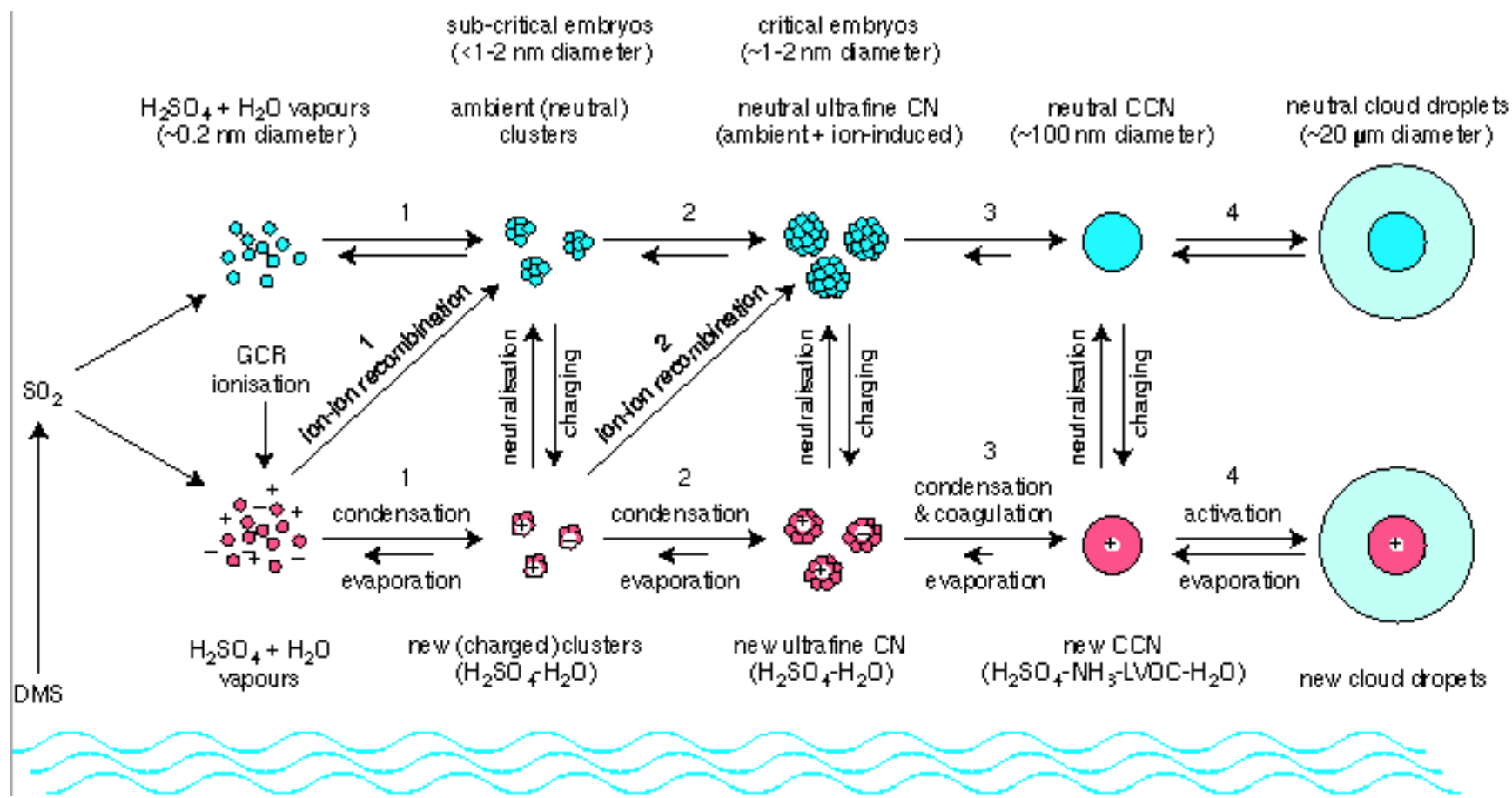,height=120mm}}
  \end{center}
  \vspace{-5mm}
  \caption{Possible influence of galactic cosmic rays (GCRs) on the
nucleation of new condensation nuclei (CN) and on their growth into cloud
condensation nuclei (CCN) in the marine boundary layer, ultimately
increasing the number of cloud droplets. The labelled steps comprise: 1)
formation of molecular clusters from condensable vapours, 2) formation
of critical embryos (ultrafine CN) from molecular clusters, 3) growth of
CN into CCN, and 4) activation of CCN into cloud droplets.  The upper
path corresponds to neutral (uncharged) aerosol particles and the lower
path to charged aerosol particles (from GCR ionisation). Charged
aerosols are expected to have an enhanced growth rate and reduced
evaporation relative to neutral aerosols.   Dimethyl sulphide (DMS) from
plankton is the major source of sulphur dioxide---the precursor of
sulphuric acid---in remote marine environments.}
  \label{fig_gcr_to_cn}    
\end{sidewaysfigure}

The steps involved in the creation of CCN from condensable vapours (in
this case, sulphuric acid) are shown in Fig.\,\ref {fig_gcr_to_cn}. 
Molecular \htwosofour-\htwoo\  clusters form and evaporate continually
(step 1) by kinetic motion.  Under suitable conditions some clusters
will reach the critical size of about 1--2 nm diameter---known as
\emph{nucleation} (step~2). Once the critical size is reached, continued
growth of the cluster becomes preferential thermodynamically.   

The nucleation of aerosols in the atmosphere involves several competing
processes which include molecular clustering, evaporation, scavenging of
condensable vapour by pre-existing particles, and sedimentation by
rainfall. In this environment, electrically charged embryos have a
competitive advantage over neutral embryos. Charged clusters provide
additional  electrostatic attractions with polar molecules, allowing
critical embryos to form with fewer molecules than for neutral
clusters.  Therefore ions can greatly enhance the rate of formation of
new particles in regions where the concentration of condensable vapours
is too low for stable neutral clusters to form at an appreciable rate,
such as frequently occurs in the marine boundary layer.
 The key parameters controlling the rate of new particle formation are
the concentrations of condensable vapours, the GCR ionisation rate, and
the surface area of pre-existing particles.   

Once formed, the ultrafine condensation nuclei (UCN) continue to grow
(step 3).  The main growth process up to diameters of about 10 nm is
molecular condensation; for larger sizes the main growth mechanism is
coagulation of existing CN.  During the growth of CN to CCN, other
vapours in the atmosphere such as ammonia, nitric acid and organic
compounds  are known to be important. The growth rate is also expected
to be enhanced by the presence of ions, becoming less significant with
increasing size of the aerosol particle.  This enhancement is largest
when the two colliding particles have opposite sign ($+ -$), but there
is also an increased rate between one charged and one neutral particle
($+ 0$ or $- 0$), when compared with two neutral particles ($0 0$). 
Finally the activation of CCN into cloud droplets (step 4) may also be
influenced by charge.  However the effect is expected to be small for
the typical electric charges found on aerosol particles under
fair-weather conditions. 

There is a continual interchange between charged and neutral particles
as small ions diffuse onto existing CN and CCN, either neutralising or
charging them in the process.  This implies that ions may potentially
affect the production  rate  of a large fraction of the CCN produced by 
gas-to-particle conversion, regardless of whether or not the original
UCN were produced via ion-mediated processes. 

It is important to note that aerosol particles and trace vapours are
continually being scavenged from the atmosphere by rainfall.  Following a
precipitation event, the air is left with a relatively low aerosol
particle number concentration and a low aerosol surface area. It is
under these conditions that the nucleation of fresh UCN is most likely to
occur.  Furthermore, for clean environments, this indicates that the
\emph{rate} at which new particles are produced and grow into CCN can
strongly affect the lifetime and radiative properties of the clouds in
these regions, since there is rarely sufficient time for a large CCN
population to form.

\subsection{Aerosol effects on cloud radiative properties}  
\label{sec_cloud_radiative_properties}

The effect that a change in the CCN number concentration has on the
radiative properties of a cloud can be quantitatively estimated as
follows \cite{twomey91,hobbs}.  Assuming the liquid water content and
depth of the cloud is fixed, then its optical thickness, $\tau$, is
given by
$\tau \propto N r^2$, where $N$ is the droplet number concentration and
$r$ the mean droplet radius. Since $N \propto r^{-3}$, this indicates 
$\tau \propto N^{1/3}$. Therefore a  change of the droplet number
concentration by
$\Delta N$ leads to a change of the optical thickness by $\Delta \tau$,
where
\begin{eqnarray}
 {{\Delta \tau } \over \tau }={1 \over 3}\cdot {{\Delta N} \over N} 
 \label{eq_dt/t}
\end{eqnarray} 
 The albedo (reflectivity), $A$, of a cloud is the fraction of  incident
radiation that is reflected into the backward hemisphere.  For the
scattering of solar radiation by clouds \cite{twomey91,hobbs},
\begin{eqnarray}
 A\approx {\tau  \over {\tau +6.7}} 
 \label{eq_a}
\end{eqnarray} Differentiating Eq.\,\ref{eq_a} and combining with 
Eq.\,\ref{eq_dt/t} gives,
\begin{eqnarray}
 {{\Delta A} \over A}=(1-A)\cdot {{\Delta N} \over N} 
 \label{eq_da/a}
\end{eqnarray}
 The rather thin stratiform clouds that cover an appreciable fraction of
the Earth's surface, and especially marine regions,  have an albedo of
about 0.5 and a droplet number concentration of about 100 \pcc\ or less
(see Fig.\,\ref{fig_ccn_concentrations}).  Equation
\,\ref{eq_da/a} shows that these clouds are very sensitive to changes in
the CCN number concentration; their reflectance changes by 0.5\% or more 
per single additional cloud droplet per cubic centimetre of air!  This in
turn indicates that GCR-induced changes in the CCN number concentration
of only a few per cent could produce significant effects on the radiative
properties of such clouds.

\section{Exp.1: Nucleation of ultrafine CN} 
\label{sec_cn_nucleation}

\subsection{Model predictions}  \label{sec_model_predictions_1}

The recent model study of Yu and Turco \cite {yu00a} provides
quantitative predictions of the ion-mediated nucleation of ultrafine
condensation nuclei (UCN) and their subsequent growth,  under the
ambient conditions measured in the Pacific boundary layer by Clarke
\emph{et al.} \cite{clarke}. Their model incorporates the effects of 
electrostatic interactions between charged and neutral molecules and
clusters.   Although the basic physics is well known, several parameters
of the model are poorly constrained by experimental data---such as the
`sticking' probability when two particles collide---and, in these cases, 
reasonable values have been estimated.  The model predictions should be
therefore be taken as representative of the magnitude of ion-mediated
effects, rather than as definitive.  Nevertheless the model studies of
Yu and Turco represent the most quantitative that presently exist on the
possible effects of GCR ionisation on aerosol formation and growth under
atmospheric conditions.

\begin{figure}[hbp]
  \begin{center}
      \makebox{\epsfig{file=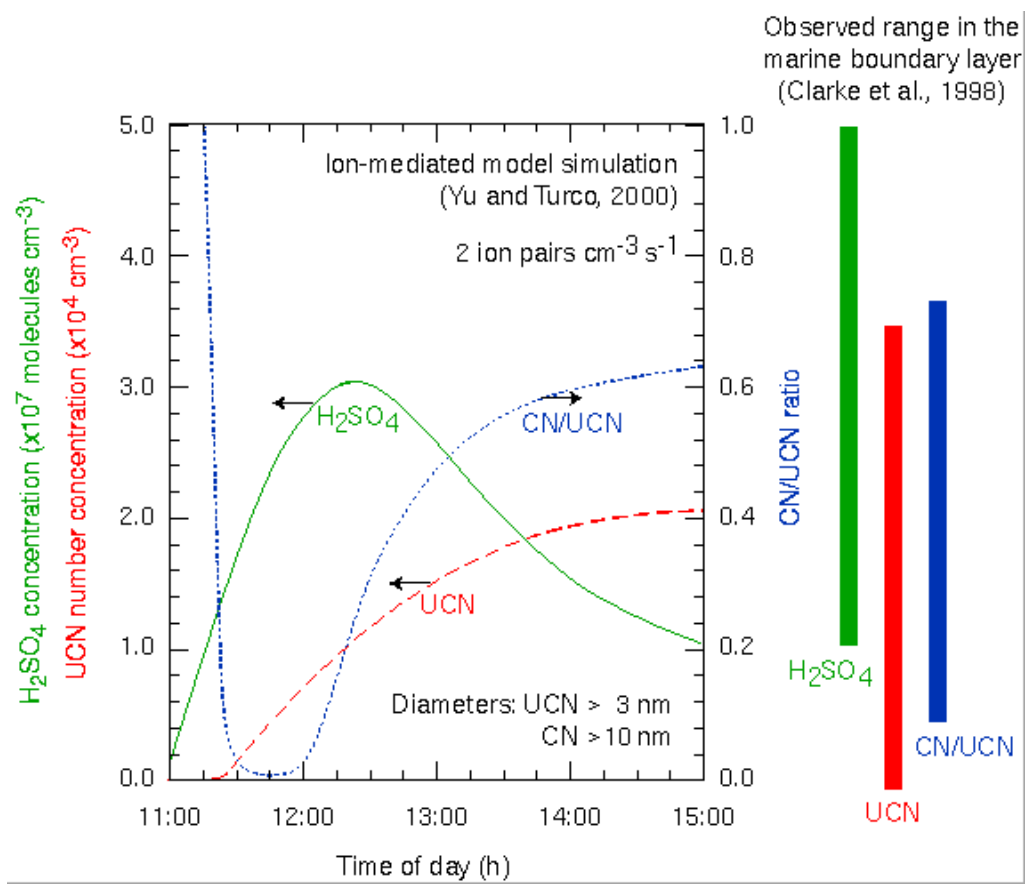,width=135mm}}
  \end{center}
  \vspace{-5mm}
  \caption{Ion-mediated model simulation by Yu and Turco
\protect{\cite{yu00a}} of the binary nucleation of new ultrafine
condensation nuclei (UCN) from \htwosofour-\htwoo\ vapour in the marine
boundary layer. An initial background  CN  concentration of  100
cm$^{-3}$ is assumed. The simulations agree well with the range of
values measured by Clarke \emph{et al.}\,\protect{\cite{clarke}},
indicated by the vertical bars on the right hand side.  Classical binary
homogeneous nucleation theory, which does not include ion-mediated
effects, predicts no nucleation under these conditions (see
Fig.\,\ref{fig_nucleation_rate}).}
  \label{fig_yu_clarke}    
\end{figure}

To compare directly with the data of Clarke \emph{et al.}, the
simulation is initialised at 11:00 local time under the
experimentally-observed conditions, namely T = 298 K, relative humidity
(RH) = 95\% and an  initial background aerosol distribution of 100 \pcc. 
Sulphuric acid vapour is generated \emph{in situ} at the observed rate of
\begin{eqnarray} 
  10^4 \cdot \sin [\pi(t-6)/12] \;\; 
  \mathrm{molecules \; cm}^{-3} \mathrm{s}^{-1}
\label{eq_h2so4_production}
\end{eqnarray}   
  where 6 $< t  <$ 18 h is the local time. The \htwosofour\ production
rate is set to zero at other times. The ion pair production rate is
2\,\pcc s$^{-1}$, corresponding to GCRs at ground level.

\vspace{5mm}

\begin{figure}[hbp]
  \begin{center}
      \makebox{\epsfig{file=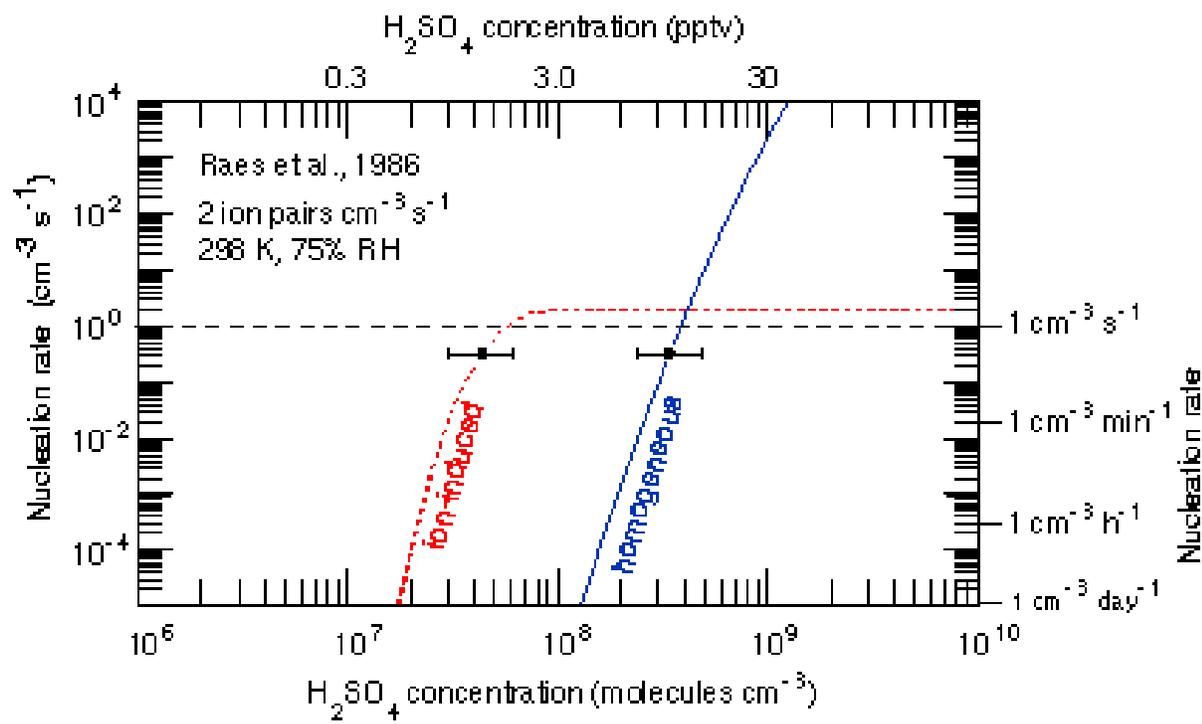,width=160mm}}
  \end{center}
  \caption{The rates of ion-induced and classical binary homogeneous
nucleation of \htwosofour-\htwoo\ aerosol particles as a function of
\htwosofour\ concentration in the marine boundary layer, estimated by
Raes \emph{et al.}
\cite{raes86}. A background ($d$\,=\,100\,nm) CN  concentration of  400
cm$^{-3}$ is included.  The expected measurement precisions for CLOUD
are indicated by points with error bars. 
 More recent estimates by Viisanen, Kulmala and  Laaksonen
\cite{kulmala90,viisanen} suggest the curve labelled `homogeneous' in
the figure  should be shifted to the right by about a factor five higher
\htwosofour\ concentrations.}
  \label{fig_nucleation_rate}    
\end{figure}

\vspace{5mm}

The results of the simulation are shown in Fig.\,\ref{fig_yu_clarke}, in
comparison with the experimental data. The model predicts negligible UCN 
($>$3 nm diameter) formation until the \htwosofour\ concentration
reaches about $10^7$
\pcc\ (at about 11:15 local time) and then, within one hour, a UCN 
population of about $10^4$
\pcc\ has formed.  The \htwosofour\ concentration falls after 12:00 due
to a decreased  production rate and to scavenging by the aerosol
particles. The newly-formed particles continue to grow, and after 12:00
an appreciable fraction exceed 10 nm diameter (labelled `CN').  The
simulation agrees well with the experimental observations of Clarke
\emph{et al.}, which are indicated by the vertical bars in 
Fig.\,\ref{fig_yu_clarke}.

Under these conditions classical binary homogeneous nucleation theory
predicts an insignificant production of new particles.  Figure
\ref{fig_nucleation_rate}, for example, shows the results of an earlier
calculation by Raes, Janssens and Van Dingenen \cite{raes86}, which 
indicates negligible new particle formation (i.e. below the so-called 
\emph{critical nucleation rate} of 1 \pccps) until the
\htwosofour\ concentration reaches about $4 \cdot 10^8$ molecules
\pccps.  Raes \emph{et al.} estimate the threshold for ion-induced
nucleation occurs at an \htwosofour\ concentration  about one order of
magnitude lower.   (Yu and Turco estimate the ion-mediated threshold to
be still lower by about a factor of four, i.e. near $10^7$
molecules\,\pccps.  This is mainly due to the higher RH and lower
background aerosol concentration in Yu and Turco's simulation.) Note
also from Fig.\,\ref{fig_nucleation_rate} that the maximum production
rate of new aerosols by ion-induced nucleation is predicted to be
limited  by the GCR intensity (2 \pccps\ at ground level).

\subsection{Experimental goals} \label{sec_goals_1}

The purpose of this experiment is to measure the effect of ionising
particle radiation on the rate of formation of ultrafine condensation
nuclei in the few-nm size range, in the presence of sulphuric acid and
water vapours.  Sulphuric acid is chosen because field measurements
indicate it is an important precursor vapour for gas-to-particle
conversion in the atmosphere. 

 The basic parameter to be measured is the
\emph{nucleation rate}, $J$ [\pccps], of UCN as a function of the
primary experimental variables, namely, sulphuric acid vapour
concentration, relative humidity, temperature, background aerosol
concentration  and  ion-pair production rate.  A representative
expectation of the variation of $J$ with sulphuric acid vapour
concentration and ion-pair production rate is shown in
Fig.\,\ref{fig_nucleation_rate}.  This figure indicates that a  large
measurement range is required  for the nucleation rate since it is a
steep function of parameters such as vapour concentration (and also
temperature).

The nucleation rate is determined by a measurement of the UCN number
concentration [\pcc] after a time $t$ [s]. In the case of the cloud
chamber, the minimum detectable particle number (after activation to
visible droplets) is about 1 particle per 10 cm$^3$.  Assuming an
exposure time of an hour (3600 s), this indicates the minimum measurable
nucleation rate is $J \sim 3 \cdot 10^{-5}$ \pccps.  The maximum
nucleation rate that can be measured is about 12 orders of magnitude
higher.  The maximum measurable droplet density in the cloud chamber is
about $10^7$~\pcc.  Depending on the variable under study, if the
effective exposure time is adjusted to about 1~s duration then a maximum
$J$ of $10^7$ \pccps\ can be measured.  If the variable under study is
ion pair  concentration then this is achieved by initiating a cloud
chamber expansion cycle 1 s after the arrival of the beam pulse.  If the
variable is sulphuric acid saturation ratio, then a short, precise
saturation pulse can be generated by a piston expansion during the
nucleation phase, followed by a small re-compression which arrests
nucleation but allows droplet growth and observation.  We have used the
latter technique extensively for nucleation studies with the Vienna
cloud chambers (see, for example, ref.\,\cite{wagner81}). 

\subsection{Experimental measurements}  \label{sec_measurements_1}

The experimental conditions for the first run are as follows:
\begin{tabbing}
\hspace{30mm} \=\hspace{65mm} \=\hspace{12mm} \=  \kill
\>\htwosofour\ molecular concentration   \>$5 \cdot 10^7$ \>\pcc\   \\
\>Relative humidity \>100  \>\% \\
\>Temperature   \>298 \>K   \\
\>Background aerosol concentration   \>0 \>\pcc\ \\
\>Pressure \>101 \>kPa \\
\>Beam intensity \>$10^3$ \>$\pi$/p pulse$^{-1}$   
\end{tabbing}

The carrier gas is pure artificial air (80:20 N$_2$:O$_2$).  The
procedure is first to adjust the cloud chamber and reactor chamber to
the desired temperature and then to fill both of them with the specified
gas mixture at the desired pressure. 

Before starting the run, thermodynamic equilibrium is established. 
During this period the clearing field is used to clear ions produced by
GCRs from the chambers.  The clearing time for small ions  is about 2 s
in the cloud chamber and 8 s in the reactor chamber.  This should be
compared with the typical ion lifetime of  $\sim 1000$\,s under
atmospheric conditions and with the time needed for an ion to attract an
\htwosofour\ molecule of $\sim 50$\,s at these concentrations. 
Therefore when the clearing field is present it reduces the ambient ion
pair concentration from GCRs by about a factor 100 below the natural
level, and also essentially excludes ion-mediated nucleation processes. 
The clearing field is therefore a very effective tool for measurements
at   ``zero GCR" ionisation. 

\begin{figure}[htbp]
  \begin{center}
      \makebox{\epsfig{file=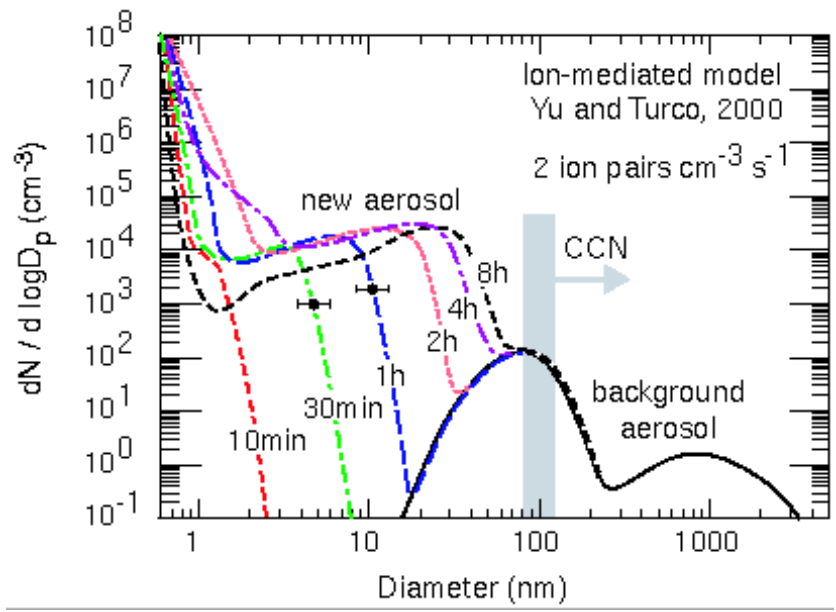,width=110mm}}
  \end{center}
  \caption{Evolution of the CN size spectra from ion-mediated nucleation
and growth under conditions typical of the marine boundary layer 
 \protect{\cite{yu00a}}. The expected measurement precisions for CLOUD
are  indicated by the  points with error bars. The solid curve indicates
the initial assumed background  aerosol  concentration of  100
cm$^{-3}$. When sulphuric acid-water CN reach diameters of about 100 nm
they become efficient CCN for seeding clouds.}
  \label{fig_cn_size_spectra}    
\end{figure}

\begin{figure}[tbp]
  \begin{center}
      \makebox{\epsfig{file=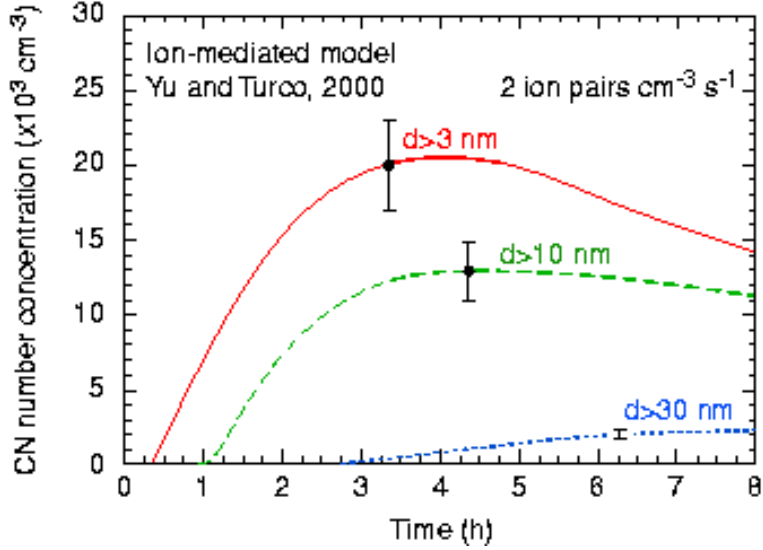,width=100mm}}
  \end{center}
  \caption{Ion-mediated nucleation and growth of CN under conditions
typical of the marine boundary layer \protect{\cite{yu00a}}.  Each curve
shows the evolution of the number of CN above an indicated  minimum
diameter,
$d$.  The expected measurement precisions for CLOUD are indicated by
points with error bars.}
  \label{fig_cn_growth_8h}    
\end{figure}

After thermodynamic equilibrium is established, the chambers are exposed
to the beam for the required period.  In the case of UCN nucleation and 
growth to $\sim 10$\,nm diameter, the characteristic times at 
atmospheric vapour concentrations are up to about one hour
(Figs.\,\ref{fig_cn_size_spectra} and \ref{fig_cn_growth_8h}), with the
first UCN appearing after about 10 minutes.  During the beam exposure
small quantities of gas are continually drawn off from the reactor
chamber via the sampling probes for analysis in the external mass
spectrometers, mobility spectrometers, and particle sizers and
counters.  In this way the evolution of the UCN nucleation and growth
can be analysed and recorded  continuously during the 1-hour beam
exposure.

In order to detect and measure UCN concentrations in the cloud chamber,
a large volume expansion ratio of about 35\% is required to create a
water vapour supersaturation of about 600\%.  Under these conditions, the
cloud chamber is sensitive to all particles above 1.2 nm diameter and
and to all \emph{charged} particles down to molecular sizes of 0.2 nm
(see Fig.\,49 in the proposal).  In the case of commercial particle
sizers the lowest detection limit is about 3 nm diameter.  Therefore,
when studying UCN nucleation, the cloud chamber will provide information
complementary to the external particle sizers.  The latter will be used
to measure the UCN size spectra versus time  (i.e. the distributions
shown in Fig.\,\ref{fig_cn_size_spectra}). The cloud chamber, in
contrast, integrates the particle count above a certain minimum size
which is set by the expansion ratio.  Once an expansion and droplet
activation cycle has taken place, the cloud chamber must be refilled
with a fresh mixture since the residual aerosols after droplet
evaporation may have been altered during the activated  phase.

During each run the following  quantities will be measured with the
detectors\footnote{The acronyms are as follows: BCS (beam counter
system),  CIMS (chemical ionisation mass spectrometer), CPC
(condensation particle counter), IOMAS (ion mass spectrometer), PIMS
(programmable ion mobility spectrometer), PITMAS (quadrupole ion trap
mass spectrometer), SMPS (scanning mobility particle sizer) and ToFMS
(time-of-flight mass spectrometer).}  indicated:
\begin{tabbing}
\hspace{30mm} \= \hspace{70mm} \=   \kill
\>\underline{\textbf{Quantity}}  \>\underline{\textbf{Detector}} 
\\[0.5ex]
\>UCN nucleation rate, $J$ [\pccps] \>Cloud chamber, CPC  \\ 
\>UCN size spectrum \>SMPS\\
\>UCN mass spectrum ($< 10^4$ amu) \>ToFMS, IOMAS \\
\>UCN chemical composition \>PITMAS \\
\>Charged UCN mobility spectrum \>PIMS   \\
\>Trace gas concentrations \>CIMS \\
\>Small ion concentration \>PIMS \\
\>Ion pair production rate \>BCS 
\end{tabbing} 
 The expected experimental precisions for the first two quantities are
indicated by the error bars in
Figs.\,\ref{fig_nucleation_rate}--\ref{fig_cn_growth_8h}.  In addition
the various thermodynamic conditions---such as temperature, pressure and
relative humidity---will be recorded.  

After completing a run with a fixed set of conditions, the `primary'
variable under study  is changed and the next run is started.  The
primary variables are those for which the nucleation rate may be most
sensitive, and which will therefore be studied in detail.  They include
the \htwosofour\ concentration, relative humidity, temperature,
background aerosol concentration, and ionisation rate.  Pressure is a
secondary variable.  The approximate ranges of study for the primary
variables are as follows:
\begin{tabbing}
\hspace{30mm} \=\hspace{65mm} \=\hspace{25mm} \=  \kill
\>\htwosofour\ concentration   \>$3 \cdot 10^6 - 10^{10}$ \>\pcc\   \\
\>Relative humidity \> $40 - 100$  \>\%  \\
\>Temperature   \>$268 - 298$ \>K   \\
\>Background aerosol concentration   \>$0 - 1000$ \>\pcc\ \\
\>Beam intensity \>$0 - 10^3$ \>$\pi$/p pulse$^{-1}$   
\end{tabbing} This series of measurements should allow CLOUD to
determine the effect of GCR ionisation on the nucleation of sulphuric
acid-water UCN under conditions corresponding to the marine boundary
layer.  This experiment is estimated to require 4 weeks beamtime (\S
\ref{sec_beam_schedule}).

\section{Exp.2: Growth of CN into CCN} 
\label{sec_cn_growth}

\subsection{Model predictions}  \label{sec_model_predictions_2}

Yu and Turco have extended their ion-mediated simulation in time  to
follow the growth of freshly formed UCN into CCN over a period of
several days \cite {yu00c}. The initial conditions are the same as
described in \S \ref{sec_model_predictions_1} except that organic
vapours are now  included in the simulation, and an initial background
aerosol distribution of 10 \pcc\ is assumed, which corresponds to clean
air, freshly scavenged by precipitation.  Although
\htwosofour\ vapour is important for the nucleation of new aerosols, the
atmospheric concentrations are generally too small to account for all of
the subsequent rate of growth of CN into CCN, and so additional vapours
are implicated.  The model assumes constant production rates
 for non-volatile organic compounds of 0.5\,$\mu$g\,m$^{-3}$\,day$^{-1}$
and   for low volatile organic compounds of
5\,$\mu$g\,m$^{-3}$\,day$^{-1}$, as suggested by field  observations.   

\begin{figure}[tbp]
  \begin{center}
      \makebox{\epsfig{file=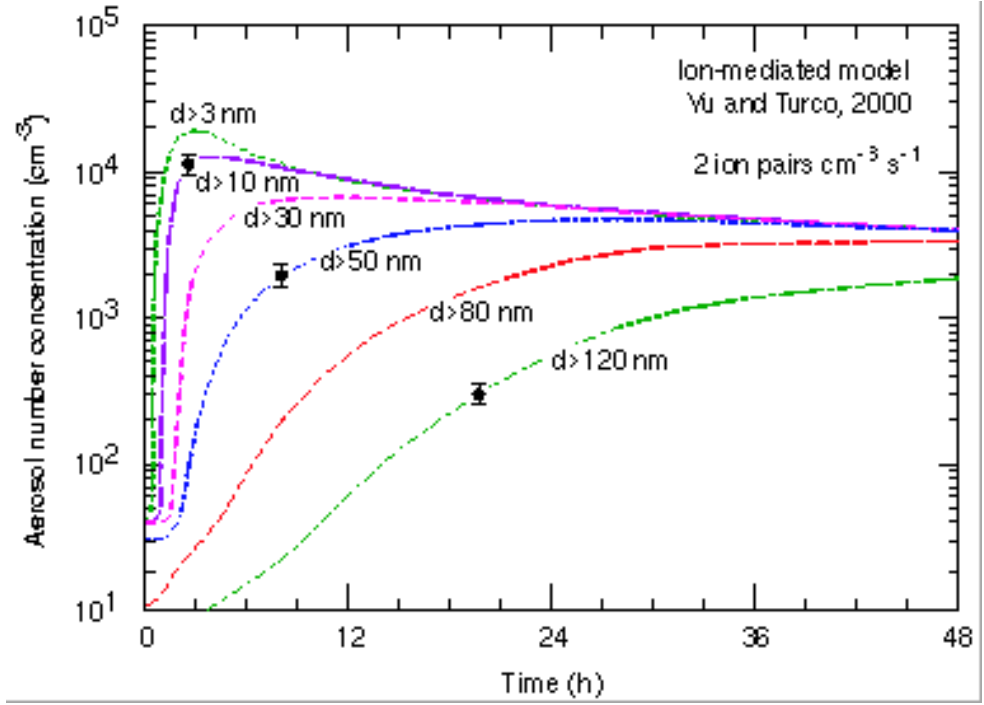,width=130mm}}
  \end{center}
  \caption{Ion-mediated model predictions of the nucleation of UCN and
growth into CCN in the presence of trace 
\htwosofour\ and  organic vapours,  under conditions typical of the
marine boundary layer 
\cite{yu00c}.  When these aerosols reach diameters of about 100\,nm they
become efficient CCN.   
 The expected measurement precisions for CLOUD are indicated by the
points with error bars.  An initial background  CN  number concentration
of  10 cm$^{-3}$ is assumed. Note that dilution (mixing) has not been
considered in these simulations, so the final aerosol concentrations are
much higher than are typically found in the marine boundary layer
($\sim$100 \pcc).}
  \label{fig_cn_growth_48h}    
\end{figure}

The results are shown in Fig.\,\ref{fig_cn_growth_48h}. The simulations
indicate a significant production of CCN (diameters above $\sim$100\,nm)
by ion-mediated processes after about 24~h.  These characteristic times
for the \emph{in situ} growth of new CN and CCN are consistent with our
field measurements \cite{kulmala00,makela}.    Note in
Fig.\,\ref{fig_cn_growth_48h} that, although the
\htwosofour\ vapour production rate peaks each day at 12:00
(Eq.\,\ref{eq_h2so4_production}), it does not lead to the formation of
new UCN after the first day because of scavenging by the existing
aerosols.

\subsection{Experimental goals} \label{sec_goals_2}

The purpose of this experiment is to measure the effect of ionising
particle radiation on the rate of growth of CN  into CCN (i.e. from
$\sim$5\,nm diameter to $\sim$100\,nm) in the presence of condensable
vapours that are  known to be important in the atmosphere: sulphuric
acid, water, ammonia and organic compounds.

\begin{figure}[tbp]
  \begin{center}
      \makebox{\epsfig{file=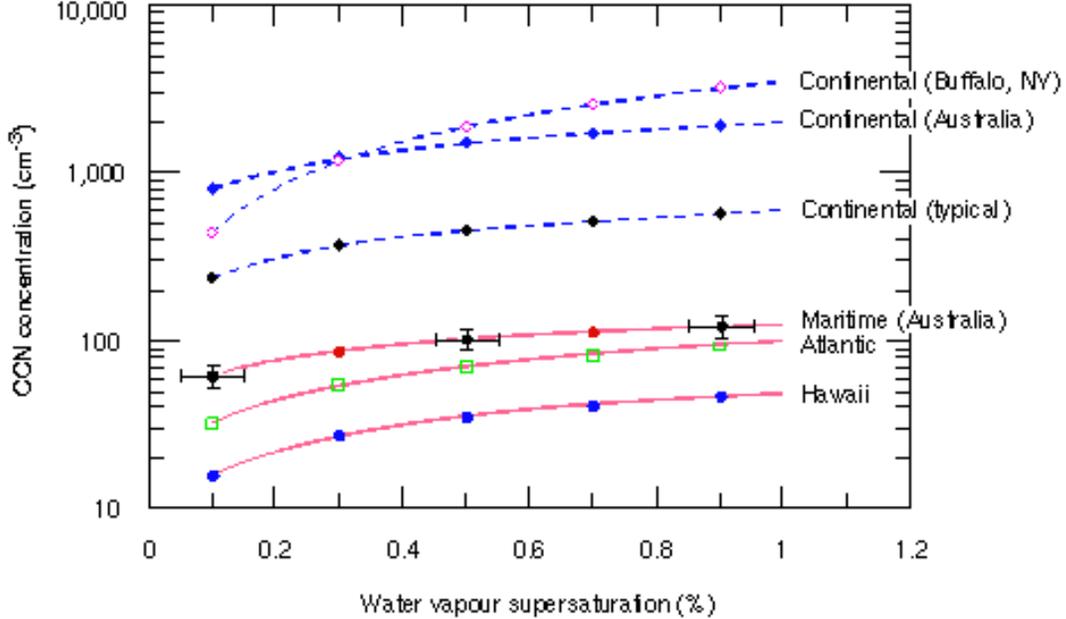,width=140mm}}
  \end{center}
  \caption{Measurements of CCN concentrations at several sites: marine
(solid curves) and continental (dashed curves) as a function of the water
vapour supersaturation \cite{seinfeld}.   The CCN concentrations are
equal to the cloud droplet concentrations at a given supersaturation. 
The expected measurement precisions for CLOUD are indicated by the
points with error bars.}
  \label{fig_ccn_concentrations}    
\end{figure}

The basic parameter to be measured is the CCN concentration, $CCN(s)$.  
The CCN are the subset of the CN population that activate into droplets
at a specified water vapour supersaturation, $s$. 
  The fraction of the CN that constitute CCN depends not only on their
size but also on their chemical composition and other properties.  
Field measurements show that $CCN(s)$ [cm$^{-3}$] can often be
parametrised as \cite{twomey59}
$$ CCN(s) = c \, s^k$$
 where s [\%] is the water vapour supersaturation, and $c$ [cm$^{-3}$]
and $k$ are empirical parameters that embed the information about the
size and chemical composition of the aerosol population.  Measurements
of $CCN(s)$ versus $s$ at several sites  are shown in
Fig.\,\ref{fig_ccn_concentrations} \cite{seinfeld}.  This figure
illustrates the low CCN population (10--100\,\pcc) that characterises
clean marine regions. The peak values of $s$ reached in the atmosphere
depend on the spectrum of CCN and on other conditions such as the
updraft velocity of the air parcel.  For marine stratiform clouds, the
typical values of
$s$ are in the range 0.1 to 0.5\%.  Even for strongly convective clouds
the peak values of $s$ are rarely above 1\% since droplet activation
arrests further increase. In the CLOUD experiment the CCN concentration
is measured with the cloud chamber operated with a small volume
expansion ratio of
$\sim 10^{-3}$, corresponding to a piston movement of $\sim$500\,$\mu$m
(see Table 5 on p.65 of the proposal).  The expected measurement
precision of $CCN(s)$ is indicated in Fig.\,\ref{fig_ccn_concentrations}.

Other important parameters to measure in this experiment are the
size-discriminated aerosol number concentrations 
(Fig.\,\ref{fig_cn_growth_48h}) and the aerosol size spectra 
(Fig.\,\ref{fig_cn_size_spectra}).  Both will be measured versus time
over the duration of each run, which lasts about 2 days.

\subsection{Experimental measurements}  \label{sec_measurements_2}

The experimental conditions for the first run are the same as for  the
previous experiment (\S \ref {sec_measurements_1}), with the addition
of  ammonia and low volatility organic compounds (LVOCs)\footnote{LVOCs
that may be important for aerosol growth in the atmosphere include
terpenes, acetone, isoprene and  the higher
$n$-alkanes.} in trace amounts:
\begin{tabbing}
\hspace{30mm} \=\hspace{65mm} \=\hspace{3mm} \=  \kill
\>\nhthree\ concentration   \>1 \>ppbv   \\
\>LVOC concentration \> 5 \>$\mu$g\,m$^{-3}$\,day$^{-1}$ 
\end{tabbing}

The procedure is similar to the previous experiment, except in the
following respects.  The beam exposure time for a single run is now 2
days.  During this time small quantities of the trace vapours are
introduced into the cloud chamber and the reactor chamber to compensate
for wall losses and for aerosol growth.    As before,  during the beam
exposure small samples of gas are continually drawn off from the reactor
chamber via the sampling probes for analysis in the external detectors. 
In this way the evolution of the CN into CCN can be measured over the
course of the beam exposure. 

After the first expansion of the cloud chamber to measure $CCN(s)$ at,
say, $s = 0.3$\%,
 the cloud chamber is cleared and re-filled from the reactor chamber. 
The volume of the cloud chamber is less than 2\% of that of  the reactor
chamber.  Therefore a total of about 5 samples for 
$CCN(0.3\%)$ measurements can be made this way over the course of the 2
day exposure period without serious impact on the contents of the
reactor chamber.  At the end of the 2-day run, several samples will be
drawn off sequentially from the reactor chamber into the cloud chamber
in order to measure the final $CCN(s)$ distribution over the range  
$0.1 < s < 1.0$\%\footnote{The cloud chamber hydraulic system can be
programmed to provide a continuous, precise slow piston expansion if
required.  This may allow a continuous scan measurement of $CCN(s)$ to
be made over the range $0.1 < s < 1.0$\% with a single chamber filling.} 
 (Fig.\,\ref{fig_ccn_concentrations}).

During each run the following quantities will be measured:
\begin{tabbing}
\hspace{30mm} \= \hspace{70mm} \=   \kill
\>\underline{\textbf{Quantity}}  \>\underline{\textbf{Detector}} 
\\[0.5ex]
\>CCN concentration, $CCN(s)$ [\pcc] \>Cloud chamber  \\
\>CN concentration \>CPC  \\
\>CN size spectrum \>SMPS\\
\>UCN mass spectrum ($< 10^4$ amu) \>ToFMS, IOMAS \\
\>CN chemical composition \>PITMAS \\
\>Charged CN mobility spectrum \>PIMS   \\
\>Trace gas concentrations \>CIMS \\
\>Small ion concentration \>PIMS \\
\>Ion pair production rate \>BCS
\end{tabbing}

As for the previous experiment, after completing a run with a fixed set
of conditions, the primary variable under study  is changed and the next
run is started.  In view of the long duration of these runs, the primary
variables in the first round experiment will be limited to trace gas
concentrations, background aerosol concentrations, and beam intensity.  
Their approximate ranges of study  are as follows:
\begin{tabbing}
\hspace{30mm} \=\hspace{65mm} \=\hspace{20mm} \=  \kill
\>LVOC concentration \> 0.5 - 50 \>$\mu$g\,m$^{-3}$\,day$^{-1}$ \\
\>Background aerosol concentration   \>0 - 1000 \>\pcc\ \\
\>Beam intensity \>$0 - 10^3$ \>$\pi$/p pulse$^{-1}$   
\end{tabbing}  
 For the run at zero beam intensity, it may be necessary to start the
experiment with a beam exposure of about 1 hour in order to generate the
initial CN population.  For the remainder of this run, the beam would be
turned off so the growth of the CN into CCN could be measured under
zero-beam conditions.

This series of measurements should allow CLOUD to determine the effect
of GCR ionisation on the growth of CN into CCN under conditions
characteristic of the marine boundary layer.  The CCN concentrations
will be parametrised as a function of water vapour supersaturation and
other variables so the results can be readily incorporated into cloud
models.  This experiment is estimated to require 8 weeks beamtime  (\S
\ref{sec_beam_schedule}).

\section{Goals of the aerosol and cloud simulations}  
\label{sec_simulations_summary}

The experimental results obtained with CLOUD will be evaluated with
aerosol and cloud simulations.  The basic physics aims of these
simulations are as follows: 
\begin{enumerate}
\item To incorporate the microphysics of ion-mediated processes into the
aerosol and cloud models.
\item To examine the sensitivity of clouds under atmospheric conditions
to variations in the GCR intensity, in the presence of other sources of
natural  variability.
\end{enumerate}

As is familiar with Monte Carlo simulations in particle physics
experiments, there will be a close feedback between the simulations and
the experimental observations to confirm a) that the simulations closely
reproduce the experimental data and b) that the underlying  microphysics
is understood.   We expect the simulations will also eventually become
important in guiding the direction of the experimental programme.   

As well as a detector simulation, this work will require a complete
simulation of the microphysical processes under study. Already existing
within our collaboration are detailed models of aerosol nucleation,
growth and activation, and also two sophisticated cloud models---for
cumulus and marine stratus clouds, respectively.  Ion-mediated
microphysics will be incorporated into these models based on the
experimental data from CLOUD.

It is important to make clear that the present modelling expertise of
our  collaboration extends as far as single clouds.   If item 2 above
reveals discernible GCR effects on single clouds, we would seek to
provide suitable parametrisations for the climate modelling community to
explore the influence of GCR effects on the global climate.

\section{Accelerator requirements} \label{sec_accelerator_requirements}

\subsection{Beam requirements} \label{sec_beam_requirements}

\begin{sidewaysfigure}[htbp]
  \begin{center}
      \makebox{\epsfig{file=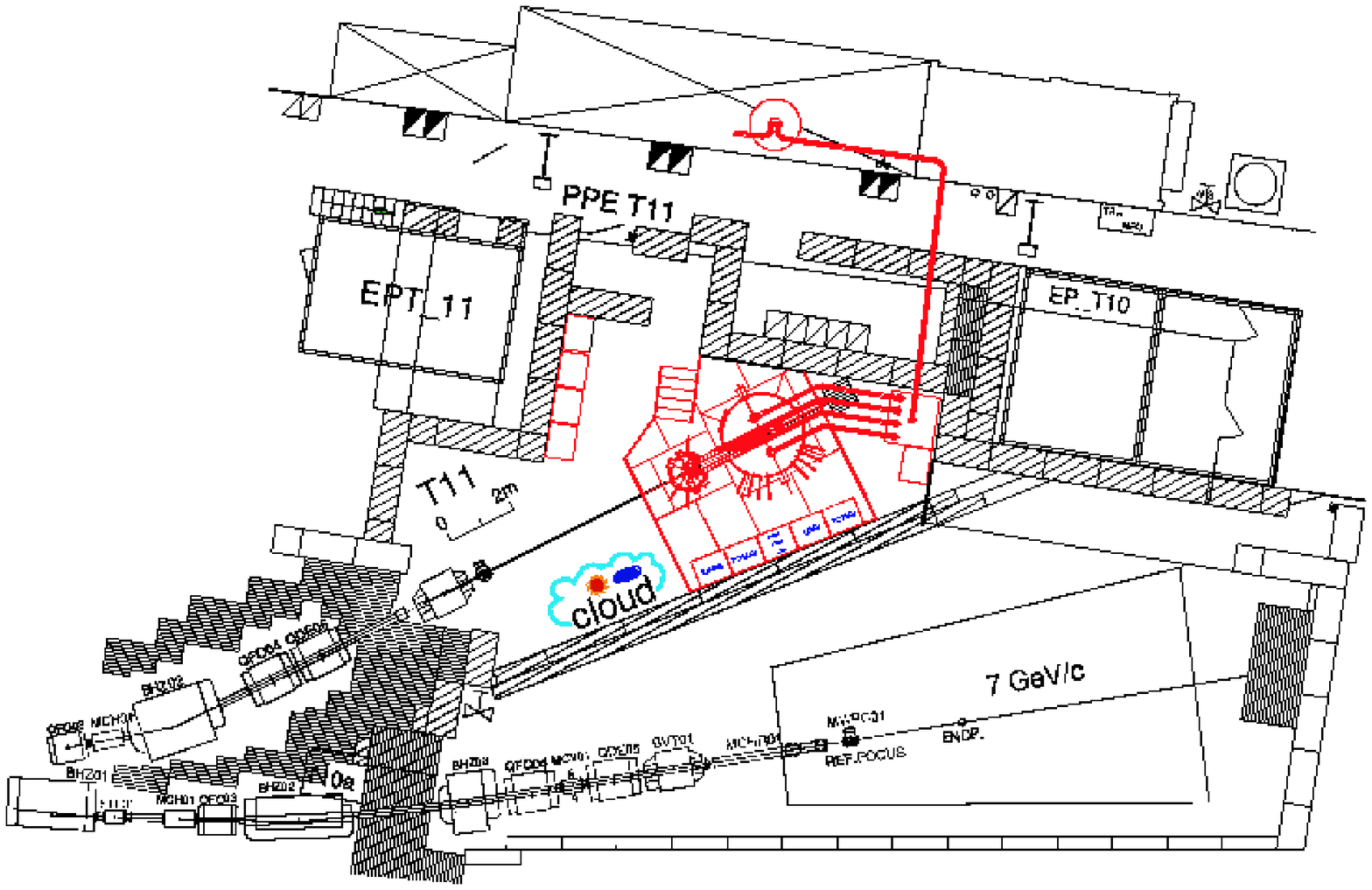,height=150mm}}
  \end{center}
  \caption{Experimental area layout of the CLOUD experiment at the CERN
PS in the T11 beamline of the East Hall.}
  \label{fig_east_hall_layout}    
\end{sidewaysfigure}

As described in the proposal (\S 7), we request to install CLOUD at the
CERN PS in the T11 beamline of the East Hall. The experimental area
layout including the new reactor chamber is shown in
Fig.\,\ref{fig_east_hall_layout}.   A dedicated counting room for CLOUD
will be installed above the current T11 counting room. The cloud chamber
and reactor chamber will be permanently located at the downstream end of
the T11 beamline, with the external detectors installed nearby.  The
upstream half of the present free space along the T11 beamline will be
left open for installing small test experiments during the beam periods
that CLOUD is not operating.  However, during its operation periods,
CLOUD would be the sole user of T11.
 
The present energy, maximum intensity and particle composition of the
T11 beam is suitable for CLOUD.  However, we have a special requirement
that the minimum beam intensity can be adjusted to about 100  particles
per pulse spread over an area of about $30 \times 30 $~cm$^2$ (i.e. near
to the natural GCR ionisation rate).  The low intensity may be reached
by closing beam collimators and perhaps adding a beam absorber.  By
suitable de-focusing of the quadrupole magnets in the T11 beamline, a
transverse size at the detector of $30 \times 30 $~cm$^2$ seems
feasible.  The transverse dimensions of the beam scintillation counters
have been increased to $40 \times 40 $~cm$^2$ to accommodate the larger
beam size (Figs.\,\ref{fig_reactor_chamber_h} and
\ref{fig_reactor_chamber_v}).

\vspace{10mm}

\begin{table*}[htbp]
\vspace{-2ex}
  \begin{center}
  \caption{Estimated beam requirements and schedule during the first
year of CLOUD operation.}
  \label{tab_2002_beam}
  \vspace{5mm}
  \begin{tabular}{| r  l | c | l|}
  \hline
  \multicolumn{2}{| c  |}{\textbf{Date}} & \textbf{No. weeks} & 
   \textbf{Activity or Experiment} \\
  \hline
  \hline
 Mar - Apr  & 2002 & 8  &  Checkout  \\
 May  & 2002  & 4 &  Exp.1: UCN nucleation \\
 Jun  & 2002  & 4 &  Exp.2: CN$\rightarrow$CCN growth  \\
 Oct  & 2002  & 4 &  Exp.2: CN$\rightarrow$CCN growth\\
 Nov  & 2002  & 4 &  Exp.3: Creation of ice nuclei   \\
 \hline 
 \multicolumn{2}{| r  |}{\textbf{Totals:}} & \textbf{8 weeks} 
    & \textbf{Checkout} \\
   & & \textbf{16 weeks} & \textbf{Data taking} \\[0.5ex]
  \hline
  \end{tabular}
  \end{center}
\end{table*}

\subsection{Beam schedule} \label{sec_beam_schedule}

The estimated beamtime required during the first year of CLOUD operation
is 24 weeks, as summarised in Table\,\ref{tab_2002_beam}.   The checkout
period includes setting up the correct operation of the hardware and
data acquisition, and carrying out technical performance tests.

\vspace{10mm}

The estimated beamtime required for each of the  two experiments
described in this document is 4 weeks and 8 weeks, respectively, as
follows:
\begin{description}

\item[Exp.1. Nucleation of UCN:] Each run is estimated to require 1
shift (8\,h) for setup and beam exposure.  The experimental variables
and the approximate number of different values that will be measured are
as follows:
\begin{tabbing}
\hspace{30mm} \=\hspace{65mm} \=4 \=  \kill
\>\htwosofour\ concentration   \>6 \>measurements   \\
\>Relative humidity \>3  \>\hspace{1.5ex} "   \\
\>Temperature   \>4 \>\hspace{1.5ex} "   \\
\>Background aerosol concentration   \>3 \>\hspace{1.5ex} "   
\end{tabbing} 
 This totals 16 runs, assuming only the variable under study is changed
in a given family of measurements.  Each of these 16 different sets of
operating conditions will be measured at 4  beam intensities: 0, 1, 3
and 10 $\times$ the natural GCR ionisation rate in the boundary layer.
(A value of `0' corresponds to the beam-off measurement.)  The total
beam time is therefore 64 shifts (3~weeks); or a calendar time of 4
weeks after allowing  for downtime, chamber cleaning, etc.

\item[Exp.2. Growth of CN into CCN:] Each run is estimated to require 6
shifts (48\,h) for setup and beam exposure.  The experimental variables
and the approximate number of different values that will be measured are
as follows:
\begin{tabbing}
\hspace{30mm} \=\hspace{65mm}  \=4 \=  \kill
\>Trace gas concentrations   \>4 \>measurements   \\
\>Background aerosol concentration   \>3  \>\hspace{1.5ex} "  
\end{tabbing} 
 This totals 7 runs, each requiring 6 shifts, to total 42 shifts.  Each
of these 7 different sets of operating conditions will be measured at 3
beam intensities: 0, 1 and 10 $\times$ the natural GCR ionisation rate. 
The total beam time is therefore 126 shifts (6~weeks); or a calendar
time of 8 weeks.

\end{description}

As regards the experimental programme for the subsequent years, 2003 and
2004, at this stage it is not possible to define the schedule precisely
since it will depend on the outcome of the initial studies as well as
developments elsewhere in the field. However we can estimate the beamtime
required for the various experiments summarised in Table\,3 (p.29) of the
proposal as follows:
\begin{tabbing}
\hspace{30mm} \= \hspace{70mm} \=   \kill
\>\underline{\textbf{Experiment}}  \>\underline{\textbf{Beamtime
[weeks]}} \\[0.5ex]
\>NO$_x$ formation \>2   \\
\>Polar stratospheric aerosols \>6   \\
\>Droplet activation \>4   \\
\>UCN nucleation II (ternary vapours) \>8   \\
\>CN $\rightarrow$ CCN growth II \>8   \\
\>Creation of ice nuclei II \>4   \\[0.5ex]
\>\textbf{Total:}  \> $\overline{\mathrm{\textbf{32 weeks}}}$    
\end{tabbing}
 Therefore during 2003 and 2004 we estimate at least 16 weeks of beamtime
will be required for CLOUD each year, which represents just over one
half of the normal yearly operation of the East Hall beamlines.

\section*{Acknowledgements}

We would like to thank Richard Turco, Fangqun Yu and Peter Zink for
stimulating and informative discussions.

\end{document}